\documentclass[11pt,a4paper]{article}
\usepackage[utf8]{inputenc}
\usepackage[T1]{fontenc}
\usepackage{amsmath}
\usepackage{amssymb}
\usepackage{cite}

\newcommand{\Dcal}{\mathcal D}
\newcommand{\dd}{\mathrm d}

\newcommand{\kfive}{\kappa_5}

\begin{document}

\title{Global Constraints for Five-Dimensional Braneworld Cosmology in Palatini and
Metric-Affine $f(\mathcal{R})$ Gravity}
\author{Paulo Michel L. T da Silva \\
S\~{a}o Paulo State Center for Technological Education (CEETEPS),\\
Faculty of Technology of Botucatu (FATEC),\\
Botucatu, S\~{a}o Paulo, Brazil\\
\texttt{paulo.silva19@cps.sp.gov.br}}
\date{06/10/2026}
\maketitle

\begin{abstract}
We investigate global consistency conditions for five-dimensional cosmological
braneworlds in Palatini and metric-affine $f(\mathcal R)$ gravity, with emphasis
on the cosmological implications of a genuinely time-dependent warped bulk. We
consider
\begin{equation*}
\mathrm{d}s^2=e^{2A(t,y)}[-\mathrm{d}t^2+a^2(t)\delta_{ij}\mathrm{d}x^i
\mathrm{d}x^j]+\mathrm{d}y^2,
\end{equation*}
and first analyze the zero-hypermomentum Palatini sector, for which the
independent connection is, up to projective freedom, the Levi-Civita
connection of $h_{AB}=F^{2/3}g_{AB}$ with $F=f_{\mathcal R}>0$. The
combination $E^i{}_i-E^4{}_4$ eliminates the scalar potential and the
algebraic $f(\mathcal R)$ term, yielding a local relation that generates a
one-parameter family of compactification constraints. The mixed $04$ field
equation provides an additional local compatibility condition coupling the
time dependence of the warp factor, the Palatini degree of freedom, and the
scalar flux. For observers confined to a $y=\mathrm{const.}$ slice, the
physical expansion rate is $\mathcal H=e^{-A}(H+\dot A)$, with
$H=\dot a/a$ the reference FLRW rate in the metric ansatz. This distinction
allows us to separate a general constant-$H$ slicing from a genuine de Sitter
induced geometry and to identify how accelerated expansion enters the global
balance. The static Einstein limit recovers the standard smooth-brane
obstruction, whereas Palatini and cosmological contributions remove the
universal sign-definite structure. We then extend the construction to genuine
metric-affine gravity in distortion form, showing how hypermomentum and
post-Riemannian curvature modify the compactification condition, and
illustrate the result with a trace-free pure-shear hypermomentum sector. The
resulting relations provide necessary consistency tests for cosmological
braneworld model building with independent connections; complete solutions
must additionally satisfy the remaining local field equations, regularity,
and compactification conditions.
\end{abstract}


\section{Introduction}

Braneworld cosmology provides a setting in which the observed four-dimensional
expansion is embedded in a higher-dimensional gravitational theory. In such
models, local bulk dynamics is not sufficient by itself: if the internal
space is compact, viable cosmological configurations must also satisfy global
integrability conditions. These constraints become especially important for
time-dependent warped geometries, because the expansion measured by observers
confined to a brane slice need not coincide with the FLRW rate appearing in a
convenient higher-dimensional coordinate ansatz.

The systematic use of global consistency conditions in braneworlds was
developed by Gibbons, Kallosh and Linde (GKL) \cite{GKL} and generalized to
arbitrary dimensions by Leblond, Myers and Winters \cite{LMW}. In the
five-dimensional Einstein--scalar case, the static sum rule contains a
non-negative scalar-gradient term and leads to the familiar obstruction to
regular periodic smooth branes under the standard assumptions. From the
cosmological point of view, the relevant question is how this balance changes
when the four-dimensional slices evolve and when the higher-dimensional
gravity sector contains additional geometric degrees of freedom.

Palatini and metric-affine $f(\mathcal R)$ gravities provide a natural
framework for addressing this question. In the Palatini sector the metric and
affine connection are varied independently, while matter is taken to be
connection independent. The connection can then be expressed through an
auxiliary conformal metric involving $F\equiv f_{\mathcal R}$. In genuine
metric-affine gravity, connection-coupled matter carries hypermomentum and may
source torsion and nonmetricity. Static Palatini thick branes and Palatini
braneworld sum rules have been studied previously
\cite{BazeiaPalatini,SilvaHoff2017}; our aim here is instead to formulate the
global consistency problem for a cosmological warped geometry and then
connect the zero-hypermomentum Palatini sector continuously to the genuine
metric-affine case.

We therefore consider 
\begin{equation}
\mathrm{d }s^2=e^{2A(t,y)}[-\mathrm{d }t^2+a^2(t)\delta_{ij}\mathrm{d }x^i%
\mathrm{d }x^j]+\mathrm{d }y^2,  \label{metric}
\end{equation}
with $A=A(t,y)$ and $\phi=\phi(t,y)$. Time-dependent warped braneworlds
have been considered in several contexts
\cite{GhoshKar,GuhaChakraborty,SilvaDutraHoffFRW}. Here the dynamical warp
factor is essential because it changes both the local mixed $04$ field
equation and the expansion rate measured by observers confined to a fixed
extra-dimensional slice. In particular, the physical brane expansion is not
simply $H=\dot a/a$ when $\dot A\neq0$.

The main results are as follows. First, in the Palatini sector we derive a
local equation from $E^i{}_i-E^4{}_4$ that eliminates the scalar potential
and the algebraic $f(\mathcal R)$ contribution and generates a one-parameter
family of global compactification constraints. Second, we retain the mixed
$04$ equation as an independent local compatibility condition, which is
necessary for any genuinely time-dependent configuration. Third, we identify
the physical expansion rate on a fixed-$y$ slice and use it to distinguish a
constant-$H$ reference slicing from an intrinsically de Sitter induced
geometry. Finally, we extend the global constraint to a genuine metric-affine
sector in terms of the distortion tensor, stress anisotropy and hypermomentum.
This makes explicit which post-Riemannian sectors can modify the sign balance
that underlies the simplest smooth-brane no-go argument.

As a consistency check, we recover the static Einstein obstruction and audit
the five-dimensional static Palatini limit. The latter comparison shows that
specializing the standard $D$-dimensional conformal Ricci relation to $D=5$
retains a mixed warp--Palatini contribution in the component identity. The
purpose of this comparison is to fix the geometric normalization used in the
present cosmological construction rather than to make a priority claim about
static Palatini sum rules.

\section{Related work and scope of the present contribution}

The ingredients entering our construction have substantial, but largely
separate, precedents. Five-dimensional thick branes in Palatini $f(R)$
gravity were constructed by Bazeia \emph{et al.}~\cite{BazeiaPalatini},
using a static warped geometry with an extra dimension of infinite extent.
Compact smooth-brane consistency conditions in the Einstein--Palatini $f(R)$
formalism were subsequently analyzed in Ref.~\cite{SilvaHoff2017}. Thus
neither Palatini thick branes nor Palatini braneworld sum rules are claimed
as new here.

Time-dependent warped braneworlds also have an established literature. In
particular, Friedmann--Robertson--Walker braneworlds generated by bulk
scalar fields with a time-dependent warp factor were studied in Ref.~\cite%
{SilvaDutraHoffFRW}, alongside other dynamical bulk constructions \cite%
{GhoshKar,GuhaChakraborty}. Our novelty claim therefore does not rest on the
use of $A(t,y)$ alone.

Post-Riemannian braneworlds have likewise been explored in geometries based
on nonmetricity. Five-dimensional thick branes in $f(Q)$ gravity were
constructed and their tensor stability analyzed in Ref.~\cite{FuZhaoXie}.
These models belong to the symmetric-teleparallel/nonmetricity class and are
conceptually distinct from the curvature-based metric-affine $f(\mathcal{R})$
theory considered below.

Finally, hypermomentum, torsion and nonmetricity are standard ingredients of
metric-affine gravity \cite{HehlMAG}. Cosmological hyperfluids and their
conservation laws have been developed systematically \cite%
{IosifidisHyperfluid}, while explicit Friedmann cosmologies have been
studied in quadratic metric-affine gravity \cite{IosifidisRavera} and, more
recently, with general hyperfluid sources \cite%
{AndreiHyperfluid,IosifidisPallikaris}. These works establish that
metric-affine cosmology with hypermomentum is itself not new.

The specific gap addressed here is the intersection of these lines of
research. We formulate compactification sum rules for a five-dimensional
cosmologically warped braneworld with an independent connection, first in
the explicit zero-hypermomentum Palatini sector and then in a genuine
metric-affine sector characterized by $\Delta_A{}^{BC}\neq0$. The
construction simultaneously retains the local mixed $04$ compatibility
equation and isolates the post-Riemannian contribution to the global sum
rule through the distortion tensor. Accordingly, the scope of the present
work is deliberately restricted to the combined consistency problem:
cosmological braneworld sum rules formulated in a framework that connects
the Palatini and genuine metric-affine sectors in the presence of
hypermomentum. No priority claim is made for any of the individual
ingredients.

\section{Independent-connection $f(\mathcal{R})$ gravity in five dimensions}

We start from 
\begin{equation}
S=\frac{1}{2\kappa_5^2}\int \mathrm{d}^5x\sqrt{-g}\,f(\mathcal{R}%
)+S_m[g,\Gamma,\phi,\Psi],  \label{actionMA}
\end{equation}
where 
\begin{equation}
\mathcal{R}=g^{AB}\mathcal{R}_{AB}(\Gamma),\qquad F\equiv f_{\mathcal{R}}.
\end{equation}
The metric stress tensor and hypermomentum are defined by 
\begin{align}
T_{AB}&=-\frac{2}{\sqrt{-g}}\frac{\delta S_m}{\delta g^{AB}}, \\
\Delta_A{}^{BC}&=-\frac{2}{\sqrt{-g}}\frac{\delta S_m}{\delta\Gamma^A{}_{BC}}%
.
\end{align}
The metric field equation is 
\begin{equation}
F\mathcal{R}_{(AB)}-\frac12 f g_{AB}=\kappa_5^2T_{AB},  \label{metricFE}
\end{equation}
and its five-dimensional trace reads 
\begin{equation}
F\mathcal{R}-\frac52 f=\kappa_5^2T.  \label{trace}
\end{equation}
Equation \eqref{metricFE} remains valid in the genuinely metric-affine
theory; hypermomentum enters the connection equation and hence the affine
curvature.

\subsection{Palatini sector}

For a canonical scalar, 
\begin{equation}
S_\phi=\int\mathrm{d}^5x\sqrt{-g}\left[-\frac12
g^{AB}\partial_A\phi\partial_B\phi-V(\phi)\right],
\end{equation}
there is no direct dependence on $\Gamma$, so $\Delta_A{}^{BC}=0$. After
fixing the irrelevant projective mode, the connection equation gives
compatibility with 
\begin{equation}
h_{AB}=F^{2/3}g_{AB}.  \label{auxmetric}
\end{equation}
Throughout the explicit Palatini construction we restrict to regular
configurations with $F=f_{\mathcal R}>0$. This guarantees a nonsingular
auxiliary conformal metric and allows the logarithmic variable to be defined
globally. Writing
\begin{equation}
u\equiv\ln F,
\end{equation}
the conformal relation in five dimensions yields 
\begin{equation}
\mathcal{R}_{AB}=R_{AB}-\nabla_A\nabla_Bu-\frac13g_{AB}\Box
u+\frac13\nabla_Au\nabla_Bu-\frac13g_{AB}(\nabla u)^2,
\label{RicciConformal}
\end{equation}
and 
\begin{equation}
\mathcal{R}=R(g)-\frac83\Box u-\frac43(\nabla u)^2.  \label{scalarConformal}
\end{equation}
All covariant derivatives on the right-hand sides are Levi-Civita
derivatives of $g_{AB}$.

\section{Cosmological warped geometry}

For the metric \eqref{metric}, define
\begin{equation}
H\equiv\frac{\dot a}{a}.
\end{equation}
Here $H$ is the expansion rate of the reference FLRW factor appearing in the
coordinate ansatz. Because the warp factor may depend on time, it is not in
general the physical Hubble rate measured on a fixed-$y$ hypersurface. For an
observer at $y=y_b$, the induced metric has proper time and physical scale
factor
\begin{equation}
\mathrm{d}\tau=e^{A(t,y_b)}\mathrm{d}t,\qquad
a_{\rm phys}(t,y_b)=e^{A(t,y_b)}a(t),
\end{equation}
so that the corresponding local expansion rate is
\begin{equation}
\boxed{\mathcal H(t,y_b)=\frac{1}{a_{\rm phys}}
\frac{\mathrm{d}a_{\rm phys}}{\mathrm{d}\tau}
=e^{-A(t,y_b)}\left[H+\dot A(t,y_b)\right].}
\label{physicalHubble}
\end{equation}
Dots and primes denote $\partial_t$ and $\partial_y$, respectively. A direct
calculation gives 
\begin{align}
R_{00}&=-3\ddot A-3H\dot A-3(\dot H+H^2)+e^{2A}(A^{\prime \prime
}+4A^{\prime 2}), \\
\frac{R_{ii}}{a^2}&=\ddot A+2\dot A^2+5H\dot A+\dot H+3H^2-e^{2A}(A^{\prime
\prime }+4A^{\prime 2}),\quad (\text{no sum}), \\
R_{04}&=-3\dot A^{\prime }, \\
R_{44}&=-4A^{\prime \prime }-4A^{\prime 2},
\end{align}
and 
\begin{equation}
R(g)=6e^{-2A}[\ddot A+\dot A^2+3H\dot A+\dot H+2H^2]-8A^{\prime \prime
}-20A^{\prime 2}.  \label{Rscalar}
\end{equation}
For $u=u(t,y)$, 
\begin{align}
(\nabla u)^2&=-e^{-2A}\dot u^2+u^{\prime 2}, \\
\Box u&=-e^{-2A}[\ddot u+(2\dot A+3H)\dot u]+u^{\prime \prime }+4A^{\prime
}u^{\prime }.
\end{align}
Substitution into Eq.~\eqref{scalarConformal} gives 
\begin{align}
\mathcal{R}={}&6e^{-2A}[\ddot A+\dot A^2+3H\dot A+\dot H+2H^2]-8A^{\prime
\prime }-20A^{\prime 2}  \notag \\
&+\frac83e^{-2A}[\ddot u+(2\dot A+3H)\dot u]-\frac83u^{\prime \prime }-\frac{%
32}{3}A^{\prime }u^{\prime }  \notag \\
&+\frac43e^{-2A}\dot u^2-\frac43u^{\prime 2}.  \label{RcalExplicit}
\end{align}

\section{Mixed compatibility equation}

The canonical scalar satisfies 
\begin{equation}
T_{04}=\dot\phi\,\phi^{\prime }.
\end{equation}
Using Eq.~\eqref{RicciConformal}, the affine mixed component is 
\begin{equation}
\mathcal{R}_{04}=-3\dot A^{\prime }-\dot u^{\prime }+A^{\prime }\dot
u+\frac13\dot u\,u^{\prime }.
\end{equation}
Since $g_{04}=0$, Eq.~\eqref{metricFE} yields the independent local
constraint 
\begin{equation}
\boxed{F\left[-3\dot A'-\dot u'+A'\dot u+\frac13\dot
u\,u'\right]=\kfive^2\dot\phi\,\phi'.}  \label{mixedconstraint}
\end{equation}
Any dynamical compact configuration must satisfy Eq.~\eqref{mixedconstraint}
in addition to the global conditions below.

\section{A local generator for cosmological Palatini sum rules}

For the scalar field, 
\begin{align}
T^i{}_i&=\frac12e^{-2A}\dot\phi^2-\frac12\phi^{\prime 2}-V, \\
T^4{}_4&=\frac12e^{-2A}\dot\phi^2+\frac12\phi^{\prime 2}-V,
\end{align}
so that 
\begin{equation}
T^i{}_i-T^4{}_4=-\phi^{\prime 2}.  \label{stressdifference}
\end{equation}
The mixed-index combination $E^i{}_i-E^4{}_4$ cancels both $f$ and $V$. The
Levi-Civita part is 
\begin{equation}
R^i{}_i-R^4{}_4=3A^{\prime \prime}+e^{-2A}\mathcal{C},
\end{equation}
where 
\begin{equation}
\mathcal{C}\equiv\ddot A+2\dot A^2+5H\dot A+\dot H+3H^2.  \label{Cdef}
\end{equation}
Equation \eqref{RicciConformal} then gives 
\begin{equation}
\mathcal{R}^i{}_i-\mathcal{R}^4{}_4=3A^{\prime \prime}+e^{-2A}[\mathcal{C}%
+(H+\dot A)\dot u]-A^{\prime }u^{\prime }+u^{\prime \prime
}-\frac13u^{\prime 2}.  \label{Rdiff}
\end{equation}
Combining Eqs.~\eqref{metricFE}, \eqref{stressdifference} and \eqref{Rdiff}, 
\begin{equation}
\boxed{3A''=-\frac{\kfive^2}{F}\phi'^2-e^{-2A}[\mathcal C+(H+\dot A)\dot
u]+A'u'-u''+\frac13u'^2.}  \label{localPalatini}
\end{equation}

\section{One-parameter family of dynamical sum rules}

Let $y$ parametrize a compact boundaryless internal dimension and assume the
relevant fields are regular and periodic. For any real $n$, 
\begin{equation}
\partial_y(e^{nA}A^{\prime})=e^{nA}(A^{\prime \prime }+nA^{\prime 2}),
\end{equation}
so 
\begin{equation}
\oint \mathrm{d }y\,e^{nA}(A^{\prime \prime }+nA^{\prime 2})=0.
\end{equation}
Multiplying Eq.~\eqref{localPalatini} by $e^{nA}$ and integrating the $%
u^{\prime \prime }$ term by parts gives 
\begin{equation}
\boxed{\begin{aligned} 0=\oint \dd
y\,e^{nA}\Bigg[&\frac{\kfive^2}{F}\phi'^2-\frac13u'^2-(n+1)A'u'-3nA'^2\\
&+e^{-2A}\{\mathcal C+(H+\dot A)\dot u\}\Bigg]. \end{aligned}}
\label{sumfamily}
\end{equation}
This is a necessary consistency condition at each cosmological time for
which the compactification assumptions hold.

\subsection{Quadratic member}

For $n=1$, 
\begin{equation}
-\frac13u^{\prime 2}-2A^{\prime }u^{\prime }-3A^{\prime
2}=-\frac13(u^{\prime }+3A^{\prime})^2,
\end{equation}
and Eq.~\eqref{sumfamily} becomes 
\begin{equation}
\boxed{\begin{aligned} 0=\oint \dd
y\,e^A\Bigg[&\frac{\kfive^2}{F}\phi'^2-\frac13\left(3A'+\frac{F'}F\right)^2%
\\ &+e^{-2A}\left\{\ddot A+2\dot A^2+5H\dot A+\dot H+3H^2+(H+\dot
A)\frac{\dot F}{F}\right\}\Bigg]. \end{aligned}}  \label{quadratic}
\end{equation}
For $F>0$ the first term is non-negative, the explicit square is
non-positive, and the cosmological contribution has no universal sign. Thus
the elementary positive-definite Einstein argument does not extend to Eq.~%
\eqref{quadratic}.

\section{Einstein and static limits}

The cleanest Einstein check follows from $n=0$. Setting $F=1$ and hence $u=0$
in Eq.~\eqref{sumfamily} gives 
\begin{equation}
\oint\mathrm{d }y\,[\kappa_5^2\phi^{\prime 2}+e^{-2A}\mathcal{C}]=0.
\end{equation}
For a static geometry, $\dot A=H=0$, so $\mathcal{C}=0$ and 
\begin{equation}
\boxed{\oint\dd y\,\phi'^2=0.}  \label{EinsteinNoGo}
\end{equation}
For a regular real scalar, Eq.~\eqref{EinsteinNoGo} forces a trivial
extra-dimensional profile. This is the expected sign-definite structure
underlying the standard smooth-brane obstruction \cite{GKL,LMW}.

In the static Palatini sector, Eq.~\eqref{localPalatini} reduces instead to 
\begin{equation}
3A^{\prime \prime }-A^{\prime }u^{\prime }+u^{\prime \prime
}-\frac13u^{\prime 2}=-\frac{\kappa_5^2}{F}\phi^{\prime 2}.
\label{staticlocal}
\end{equation}
For $n=0$ and periodic fields, 
\begin{equation}
\boxed{\kfive^2\oint\dd y\,\frac{\phi'^2}{F}=\oint\dd
y\left[A'\frac{F'}F+\frac13\left(\frac{F'}F\right)^2\right].}
\label{staticdirect}
\end{equation}
Equation \eqref{staticdirect} is the direct component result following from
Eqs.~\eqref{auxmetric}--\eqref{RicciConformal}. Its comparison with the
weighted static Palatini construction of Ref.~\cite{SilvaHoff2017} deserves
separate attention.

\section{Constant-$H$ slicing and the static-bulk de Sitter limit}

\label{sec:dS} A useful specialization of the dynamical family is obtained by
taking the reference FLRW factor to have constant coordinate expansion rate,
\begin{equation}
a(t)=a_0e^{H_0t},\qquad H=H_0,\qquad \dot H=0.
\end{equation}
For a general time-dependent warp factor this assumption should not be
identified with an intrinsically de Sitter induced geometry. Indeed, the
physical expansion rate on a fixed-$y$ slice is, from
Eq.~\eqref{physicalHubble},
\begin{equation}
\mathcal H(t,y)=e^{-A(t,y)}\left[H_0+\dot A(t,y)\right],
\label{physicalHubbleConstH}
\end{equation}
and is generically time dependent. Likewise, the intrinsic four-dimensional
Ricci scalar of a $y=\mathrm{const.}$ slice is
\begin{equation}
{}^{(4)}R=6e^{-2A}
\left[\ddot A+\dot A^2+3H_0\dot A+2H_0^2\right],
\label{R4slice}
\end{equation}
which need not be constant when $\dot A\neq0$. We therefore refer to this
case as a constant-$H$ reference slicing.

The cosmological combination entering the sum rules becomes
\begin{equation}
\mathcal{C}_{H_0}=\ddot A+2\dot A^2+5H_0\dot A+3H_0^2.
\label{Cds}
\end{equation}
Consequently the distinguished $n=1$ Palatini condition reads
\begin{equation}
\boxed{ 0=\oint \dd y\,e^A\left[ \frac{\kfive^2}{F}\phi'^2
-\frac13\left(3A'+\frac{F'}{F}\right)^2
+e^{-2A}\left( \ddot A+2\dot A^2+5H_0\dot A+3H_0^2
+(H_0+\dot A)\frac{\dot F}{F} \right) \right].}
\label{dSgeneral}
\end{equation}
This expression isolates the effect of a constant reference expansion rate
without imposing additional conditions on the induced four-dimensional
geometry.

A genuine and particularly transparent de Sitter specialization is obtained
for a time-independent bulk profile,
\begin{equation}
A=A(y),\qquad F=F(y),\qquad \phi=\phi(y),
\end{equation}
combined with $a(t)=a_0e^{H_0t}$. On each fixed-$y$ slice the conformal factor
is then time independent, and the induced geometry is de Sitter up to the
constant rescaling of proper time. The physical Hubble rate on that slice is
\begin{equation}
\mathcal H(y)=e^{-A(y)}H_0.
\end{equation}
Equation~\eqref{dSgeneral} reduces to
\begin{equation}
\boxed{ 0=\oint \dd y\,e^A\left[ \frac{\kfive^2}{F}\phi'^2
-\frac13\left(3A'+\frac{F'}{F}\right)^2 +3H_0^2e^{-2A} \right].}
\label{dSstaticbulk}
\end{equation}
The mixed $04$ equation is identically satisfied for this restricted
static-bulk ansatz. Equation~\eqref{dSstaticbulk} therefore provides a clean
necessary consistency test for a compact smooth brane with de Sitter
four-dimensional slices: the positive curvature contribution competes with
the scalar-gradient and Palatini/warp quadratic sectors. It is not an
existence theorem.

In the Einstein limit $F\to1$, Eq.~\eqref{dSstaticbulk} becomes
\begin{equation}
0=\oint \mathrm{d}y\,e^A\left[ \kappa_5^2\phi^{\prime 2}-3A^{\prime 2}
+3H_0^2e^{-2A} \right],
\end{equation}
showing explicitly that the flat-slice static positivity argument cannot be
transferred unchanged to de Sitter slicing once the intrinsic curvature term
is retained.

\section{Static-limit audit and the dimensional conformal identity}

\label{sec:static-audit} Previous static Palatini braneworld sum rules are
conventionally organized using a total derivative of the form $%
\nabla\cdot(W^\alpha\nabla W)$, with $W=e^A$ \cite{SilvaHoff2017}. Since 
\begin{equation}
\partial_y(W^\alpha W')=e^{(\alpha+1)A}[A^{\prime \prime}+(\alpha+1)A^{\prime 2}],
\end{equation}
the weighting parameters are related by 
\begin{equation}
\boxed{n=\alpha+1.}
\end{equation}
Thus the commonly discussed choice $\alpha=-1$ corresponds to $n=0$, not $n=1
$.

A more fundamental issue emerges when the conformal Ricci transformation is
audited in arbitrary dimension. Let 
\begin{equation}
q\equiv F^{2/(D-2)},\qquad h_{AB}=q\,g_{AB}.
\end{equation}
The standard $D$-dimensional conformal transformation gives 
\begin{align}
\mathcal{R}_{AB}={}&R_{AB} -\frac{D-2}{2q}\nabla_A\nabla_Bq -\frac{1}{2q}%
g_{AB}\Box q  \notag \\
&+\frac{3(D-2)}{4q^2}(\nabla_Aq)(\nabla_Bq) +\frac{4-D}{4q^2}g_{AB}(\nabla
q)^2.  \label{RicciQgeneral}
\end{align}
Equivalently, with $u=\ln F$, Eq.~\eqref{RicciQgeneral} reduces identically
to Eq.~\eqref{RicciConformal}.

The conformal Ricci expression used in Eq.~(15) of Ref.~\cite{SilvaHoff2017}%
, when translated to the same variable $q$, has the structure 
\begin{equation}
\mathcal{R}_{AB}=R_{AB} -\frac{1}{q}\nabla_A\nabla_Bq -\frac{1}{2q}%
g_{AB}\Box q +\frac{D-1}{2q^2}(\nabla_Aq)(\nabla_Bq),
\label{Ricci2017structure}
\end{equation}
with no $g_{AB}(\nabla q)^2$ term. Comparing Eqs.~\eqref{RicciQgeneral} and %
\eqref{Ricci2017structure} shows that their coefficients agree
simultaneously only for $D=4$. In particular, for $D=5$ Eq.~%
\eqref{RicciQgeneral} contains the coefficients 
\begin{equation}
-\frac{3}{2q},\qquad -\frac{1}{2q},\qquad \frac{9}{4q^2},\qquad -\frac{1}{%
4q^2},
\end{equation}
for the Hessian, metric Laplacian, gradient-square tensor and metric
gradient-square terms, respectively, whereas Eq.~\eqref{Ricci2017structure}
yields a different five-dimensional reduction.

This difference in dimensional specialization accounts for the apparent discrepancy in the static
sum rule. Ref.~\cite{SilvaHoff2017} reports, for $D=5$ and $\alpha=-1$, the
definite-sign condition 
\begin{equation}
\oint\mathrm{d }y\,\frac{\Phi^{\prime 2}}{F} +\frac{1}{27\pi G_5}\oint%
\mathrm{d }y\,[(\ln|F|)^{\prime}]^2=0,  \label{oldPalatiniRule}
\end{equation}
whereas the direct five-dimensional component calculation based on Eq.~%
\eqref{RicciQgeneral} gives Eq.~\eqref{staticdirect}, including the mixed
term $A^{\prime }F^{\prime }/F$.

Accordingly, the difference is not merely a choice of weighting parameter or
normalization convention. It arises at the stage where the conformal Ricci
relation is specialized to five dimensions before the partial-trace sum rule
is constructed. In the present work we
therefore use the standard $D$-dimensional conformal identity %
\eqref{RicciQgeneral} as the geometric starting point and retain Eq.~%
\eqref{staticdirect} as the corresponding five-dimensional static limit. The
earlier result \eqref{oldPalatiniRule} is quoted for comparison but is not
imposed as a consistency requirement on the present dynamical system.

This correction does not by itself establish the existence of smooth compact
Palatini branes. It changes the necessary integrated condition and removes
the basis for importing the previous definite-sign conclusion directly into
the five-dimensional system. Existence still requires the complete local
field equations, regularity, compactification conditions, and, in the
cosmological case, the mixed constraint \eqref{mixedconstraint}.

\section{Genuine metric-affine extension}

We now allow matter to couple directly to the affine connection, 
\begin{equation}
\Delta_A{}^{BC}\neq0,
\end{equation}
and decompose 
\begin{equation}
S_m=S_\phi[g,\phi]+S_{\mathrm{MA}}[g,\Gamma,\Psi],\qquad
T_{AB}=T^{(\phi)}_{AB}+\Theta_{AB}.
\end{equation}
The independent connection is written as 
\begin{equation}
\boxed{\Gamma^A{}_{BC}=\{^A{}_{BC}\}_g+N^A{}_{BC},}  \label{distortion}
\end{equation}
where $N^A{}_{BC}$ is the distortion. It may be decomposed into contortion
and disformation, 
\begin{equation}
N^A{}_{BC}=K^A{}_{BC}+L^A{}_{BC},
\end{equation}
associated respectively with torsion and nonmetricity, 
\begin{equation}
S^A{}_{BC}=\Gamma^A{}_{[BC]},\qquad Q_{ABC}=-\nabla_A^\Gamma g_{BC}.
\end{equation}

The curvature difference generated by Eq.~\eqref{distortion} is 
\begin{equation}
\mathcal{R}^A{}_{BCD}=R^A{}_{BCD}+2%
\nabla_{[C}N^A{}_{D]B}+2N^A{}_{[C|E|}N^E{}_{D]B},
\end{equation}
and hence 
\begin{equation}
\mathcal{R}_{BD}=R_{BD}+\nabla_A
N^A{}_{DB}-\nabla_DN^A{}_{AB}+N^A{}_{AE}N^E{}_{DB}-N^A{}_{DE}N^E{}_{AB}.
\label{RicciDist}
\end{equation}
Define 
\begin{equation}
\Dcal_{BD}[N]\equiv\mathcal{R}_{BD}(\Gamma)-R_{BD}(g),
\end{equation}
and the anisotropic post-Riemannian combination 
\begin{equation}
\boxed{\Xi_N\equiv\Dcal^i{}_i-\Dcal^4{}_4.}  \label{XiDef}
\end{equation}
Similarly, define the stress anisotropy of the connection-coupled matter
sector, 
\begin{equation}
\boxed{\Pi_\Delta\equiv\Theta^i{}_i-\Theta^4{}_4.}  \label{PiDef}
\end{equation}
Then 
\begin{equation}
T^i{}_i-T^4{}_4=-\phi^{\prime 2}+\Pi_\Delta.
\end{equation}
Using the same $E^i{}_i-E^4{}_4$ combination as before gives the local
metric-affine consistency equation 
\begin{equation}
\boxed{3A''=-\frac{\kfive^2}{F}(\phi'^2-\Pi_\Delta)-e^{-2A}\mathcal C-\Xi_N.}
\label{localMA}
\end{equation}

\section{General metric-affine sum rule}

Multiplying Eq.~\eqref{localMA} by $e^{nA}$ and using the compactness
identity gives 
\begin{equation}
\boxed{\begin{aligned} 0=\oint\dd
y\,e^{nA}\left[\frac{\kfive^2}{F}(\phi'^2-\Pi_\Delta)+e^{-2A}\mathcal
C+\Xi_N-3nA'^2\right]. \end{aligned}}  \label{MASumRule}
\end{equation}
Equation \eqref{MASumRule} is the general distortion-form consistency
condition following from the metric field equations in the genuine
metric-affine sector. Since $N=K+L$, one may
organize 
\begin{equation}
\Xi_N=\Xi_S+\Xi_Q+\Xi_{SQ},
\end{equation}
where the three terms denote pure torsion, pure nonmetricity and mixed
contributions. Schematically, 
\begin{equation}
\Xi_N\sim\nabla K+\nabla L+K^2+L^2+KL.
\end{equation}
The connection equation relates these quantities to the hypermomentum, 
\begin{equation}
P_A{}^{BC}[F,g,\Gamma]=\kappa_5^2\,\Delta_A{}^{BC},
\label{schematicconnection}
\end{equation}
with convention-dependent density factors understood. Hypermomentum can be
decomposed into spin, dilation and shear sectors \cite{HehlMAG}. Without a specified connection-dependent matter Lagrangian, the
hypermomentum may itself depend on the affine connection, so the relation
between $N$ and $\Delta$ need not constitute a closed constitutive system.
For this reason Eq.~\eqref{MASumRule} is deliberately retained in distortion
form.

\subsection{Projective consistency}

A pure $f(\mathcal{R})$ gravitational sector possesses projective freedom.
In the Palatini case this can be gauge-fixed without affecting the scalar
matter considered above. For $\Delta_A{}^{BC}\neq0$, however, the matter
coupling must be compatible with the projective structure, or the projective
invariance must be broken by an explicitly specified term. This is why we do
not impose an arbitrary nonzero dilation current in Eq.~\eqref{MASumRule}. A
model-specific hypermomentum specialization is a separate physical
assumption, not a consequence of the sum rule itself.

\section{A trace-free pure-shear hypermomentum illustration}

\label{sec:shear} To illustrate the metric-affine extension, we now specify
a trace-free hypermomentum ansatz consistent with the projective trace
constraint of a pure Ricci-scalar $f(\mathcal{R})$ gravitational action.
Projective invariance requires the dilation trace of the matter
hypermomentum to vanish; this restriction is standard in metric-affine
cosmology and prevents an arbitrary dilation current from being coupled
consistently without an additional projective-symmetry-breaking sector \cite%
{IosifidisHyperfluid}. Pure shear hypermomentum is therefore a natural
minimal choice; it has also been used explicitly in metric-affine $%
f(R)=R+\beta R^2$ cosmology \cite{IosifidisShear}.

Let $n^A$ denote the unit spacelike normal to the $y=\mathrm{const.}$
hypersurfaces, $n_An^A=1$. A simple five-dimensional, spinless and
trace-free shear ansatz is 
\begin{equation}
\boxed{ \Delta_A{}^{BC}=\sigma(t,y)\left[ \delta_A^{(B}n^{C)}+\frac12
g^{BC}n_A-\frac72 n_A n^B n^C \right].}  \label{shearansatz}
\end{equation}
It is symmetric in $(BC)$ and satisfies both 
\begin{equation}
g_{BC}\Delta_A{}^{BC}=0, \qquad \Delta_B{}^{BC}=0,  \label{sheartraces}
\end{equation}
so that it carries neither a spin current nor the projectively forbidden
dilation trace. The remaining scalar amplitude $\sigma(t,y)$ parametrizes a
shear-type source compatible with the three-dimensional spatial isotropy of
the brane while allowing the distinguished extra-dimensional direction. At
this stage Eq.~\eqref{shearansatz} is a kinematic ansatz, not a
hypermomentum derived from a microscopic matter Lagrangian.

For the pure $f(\mathcal{R})$ theory, the connection equation is linear in
the covariant derivative of the density $\sqrt{-g}Fg^{AB}$ and algebraic in
the hypermomentum. After fixing the unphysical projective mode, the connection equation may be
organized formally as a Palatini contribution plus a shear-sourced response, 
\begin{equation}
N^A{}_{BC}=N^A{}_{BC}[F]+N^A{}_{BC}[\sigma],  \label{NsplitShear}
\end{equation}
where this decomposition is understood as a formal organization after
projective gauge fixing. The second term need not be an independent linear
functional of $\sigma$ once a microscopic matter model is specified, because
the hypermomentum may itself depend on the affine connection. Here $N[F]$
denotes the Palatini conformal part and $N[\sigma]$ the additional
shear-sourced response. Substitution into the curvature difference gives 
\begin{equation}
\Xi_N=\Xi_F+\Xi_\sigma+\Xi_{F\sigma},  \label{XiShearSplit}
\end{equation}
so that the global condition becomes 
\begin{equation}
\boxed{ 0=\oint \dd y\,e^{nA}\left[ \frac{\kfive^2}{F}(\phi'^2-\Pi_\sigma)
+e^{-2A}\mathcal C +\Xi_F+\Xi_\sigma+\Xi_{F\sigma} -3nA'^2 \right].}
\label{shearsumrule}
\end{equation}
Here $\Pi_\sigma$ is the stress anisotropy of the shear-carrying matter
sector. Equation~\eqref{shearsumrule} is more restrictive than the
completely arbitrary-hypermomentum expression because the source has now
been restricted to the trace-free class \eqref{shearansatz}. A fully
microscopic closure requires a definite matter Lagrangian producing $\sigma$
and its stress tensor; only then are $\Pi_\sigma$ and the connection
response $N[\sigma]$ uniquely related. We deliberately do not invent such a
constitutive relation. This separation keeps the result model independent
while identifying a concrete, symmetry-compatible source sector for the next
stage of the construction.

The distinction is physically important. Recent metric-affine cosmology
shows that homogeneous and isotropic hypermomentum decomposes into dilation,
shear and spin sectors and that shear sources can modify cosmological
evolution \cite{AndreiHyperfluid,IosifidisShear}. In the present braneworld
setting, Eq.~\eqref{shearsumrule} shows where the corresponding
five-dimensional shear contribution enters the compactification consistency
condition: both through the matter anisotropy $\Pi_\sigma$ and through the
post-Riemannian curvature terms $\Xi_\sigma+\Xi_{F\sigma}$.

\section{Hierarchy of consistency conditions}

The construction is most naturally viewed as a hierarchy of sectors and
limits rather than as a strict set-theoretic inclusion of theories:
\begin{equation}
\boxed{
\text{general metric-affine sector}
\ \longrightarrow\
\text{zero-hypermomentum Palatini sector}
\ \longrightarrow\
\text{Einstein limit}.
}
\end{equation}
The Einstein limit has $F=1$, $N=0$ and $\Delta=0$. In the Palatini sector,
$F$ may vary while $\Delta=0$, and the connection is determined, up to
projective freedom, by the auxiliary metric \eqref{auxmetric}. In the genuine
metric-affine sector, $\Delta\neq0$ and the distortion contains additional
post-Riemannian contributions sourced by connection-coupled matter.

The physical content of the full global balance may be summarized
schematically as 
\begin{equation}
\boxed{\mathcal I_\phi+\mathcal I_{\rm Palatini}+\mathcal I_{\rm
cosm}+\mathcal I_{\rm hypermomentum}+\mathcal I_{\rm
torsion/nonmetricity}=0.}
\end{equation}
The additional terms can change the sign structure that drives the simplest
Einstein no-go argument. Their presence alone, however, does not establish a
regular compact solution.

\section{Consistency statement}

The preceding derivation may be summarized as follows.

\medskip \noindent\textbf{Proposition.} \emph{Consider a regular
five-dimensional spacetime of the form \eqref{metric}, with compact
boundaryless extra dimension, governed by metric-affine $f(\mathcal{R})$
gravity. Let the matter sector consist of a canonical scalar plus a possible
connection-coupled sector with stress anisotropy \eqref{PiDef} and
hypermomentum $\Delta_A{}^{BC}$. Then the metric field equations imply the
family of necessary global conditions \eqref{MASumRule}, with $\mathcal{C}$
given by Eq.~\eqref{Cdef} and $\Xi_N$ by Eq.~\eqref{XiDef}. For $\Delta=0$,
the connection reduces to the Palatini sector and the explicit relations %
\eqref{sumfamily} and \eqref{mixedconstraint} follow. In the further static
Einstein limit, the sign-definite condition \eqref{EinsteinNoGo} is
recovered.}

\medskip The proposition is a consistency statement, not a sufficiency
theorem. Existence requires solving the remaining field equations, the
scalar equation, the trace equation, the mixed constraint, the connection
equation, and the compactification/regularity conditions simultaneously.

\section{Discussion}

From the cosmological perspective, the central point is that the quantity
$H=\dot a/a$ entering the higher-dimensional ansatz is not, in general, the
expansion rate measured by observers confined to the brane. The latter is
$\mathcal H=e^{-A}(H+\dot A)$, so a time-dependent warp factor contributes
directly to the intrinsic expansion history of a fixed-$y$ slice. This
observation is important when assessing accelerated solutions: a constant
reference value of $H$ need not imply an intrinsically de Sitter brane unless
the warp dynamics is compatible with constant $\mathcal H$ and the induced
curvature.

The time dependence of the warp factor changes the logic of the braneworld
consistency problem. In a static construction, the sum rule may be discussed
entirely in terms of the internal coordinate. For $A(t,y)$, the integrated
relation is accompanied by Eq.~\eqref{mixedconstraint}, which couples $\dot
A^{\prime }$, $\dot F$, $F^{\prime }$ and the scalar flux $%
\dot\phi\phi^{\prime }$. A configuration that satisfies a global integral
identity but violates this local equation is not a solution of the theory.

For physical interpretation, one may identify the effective brane with a
localized region of the smooth bulk centered around some $y=y_b$ (for
example, a maximum of the scalar energy density). Cosmological observables on
that slice are then built from the induced metric, and in particular from the
physical rate $\mathcal H(t,y_b)$ in Eq.~\eqref{physicalHubble}, rather than
from the reference quantity $H=\dot a/a$ alone. This distinction becomes
essential whenever $\dot A\neq0$.

The Palatini sector also illustrates why statements about ``evading'' a
no-go theorem must be made carefully. The appearance of terms with different
signs means that the simplest positivity proof is no longer universal; it
does not imply that every $F(t,y)$ admits a smooth compact brane. The static
audit in Sec.~\ref{sec:static-audit} traces the discrepancy with the earlier definite-sign condition to the
five-dimensional specialization of the conformal Ricci relation, rather than
to the weighting parameter itself.

The literature comparison in Sec.~2 is also important for interpreting the
scope of the result. Static Palatini branes, Palatini sum rules,
time-dependent FRW branes, nonmetricity thick branes and metric-affine
hyperfluid cosmologies already exist as separate research directions \cite%
{BazeiaPalatini,SilvaHoff2017,SilvaDutraHoffFRW,FuZhaoXie,IosifidisHyperfluid,AndreiHyperfluid}%
. The contribution of the present framework is their previously uncombined
consistency problem: a cosmological braneworld sum-rule construction that
can be followed continuously from the zero-hypermomentum Palatini sector to
a genuine metric-affine sector.

The genuine metric-affine extension introduces a second layer of model
dependence. Hypermomentum is not an arbitrary external source: it is
generated by a specified connection-coupled matter action. Once such a
sector is chosen, Eq.~\eqref{schematicconnection} determines the distortion,
allowing $\Xi_N$ to be reduced to explicit torsion, nonmetricity and
hypermomentum invariants. Equation \eqref{MASumRule} is therefore best
regarded as a general distortion-form starting point for model-specific
metric-affine consistency and no-go analyses.

\section{Conclusions}

We have constructed a unified set of cosmological braneworld consistency
conditions in five-dimensional independent-connection $f(\mathcal{R})$
gravity. The analysis begins with a general time-dependent cosmological warp factor $%
A(t,y)$ and a bulk scalar $\phi(t,y)$. In the Palatini sector, $\Delta=0$,
the auxiliary metric $h_{AB}=F^{2/3}g_{AB}$ yields explicit affine-curvature
corrections. The combination $E^i{}_i-E^4{}_4$ eliminates the scalar
potential and the algebraic $f(\mathcal{R})$ term, producing the local
equation \eqref{localPalatini} and the one-parameter family \eqref{sumfamily}%
. The $n=1$ member takes the compact quadratic form \eqref{quadratic}, while
the mixed $04$ equation \eqref{mixedconstraint} supplies an independent
local compatibility condition.

The static Einstein limit reproduces the usual positive-gradient
obstruction. In the Palatini case, derivatives of $F$ modify the global
balance. A direct static-limit audit reveals a mixed $A^{\prime }F^{\prime
}/F$ term and traces its disagreement with the earlier definite-sign formula
to the conformal Ricci reduction used in that derivation. The present
results therefore use the standard $D$-dimensional conformal identity
specialized to $D=5$.

We then generalized the construction to genuine metric-affine gravity. The
decomposition $\Gamma=\{g\}+N$ leads to the universal consistency condition %
\eqref{MASumRule}, where $\Pi_\Delta$ encodes the anisotropic stress of
connection-coupled matter and $\Xi_N$ contains the torsion/nonmetricity
contribution to the affine curvature. The resulting hierarchy 
\begin{equation*}
\text{Einstein}\longrightarrow\text{Palatini }f(\mathcal{R})\longrightarrow%
\text{metric-affine }f(\mathcal{R}) 
\end{equation*}
provides a systematic route for determining precisely which geometric and
matter sectors modify the conventional smooth-brane obstruction.

We have additionally exhibited two cosmologically relevant specializations.
The constant-$H$ analysis separates the coordinate FLRW rate from the
physical expansion of a brane slice and shows explicitly how accelerated
expansion enters the global balance; the static-bulk limit yields a genuine
de Sitter slicing with a positive curvature contribution. The pure-shear
hypermomentum example illustrates how a connection-coupled source can enter
the same compactification condition without imposing a microscopic closure.
A natural next step is to choose a concrete $f(\mathcal R)$ model, such as a
Hu--Sawicki-type realization, and solve the complete local and global system
numerically. Such constructions would permit a direct assessment of which
branches of independent-connection gravity admit regular compact braneworlds
with viable cosmological expansion histories.

\section*{Acknowledgments}

The author acknowledges CAPES (Coordena\c{c}\~ao de Aperfei\c{c}oamento de Pessoal de N\'ivel Superior, Brazil).

\end{document}